\documentclass[review]{elsarticle}

\usepackage{lineno,hyperref}
\modulolinenumbers[50]

\usepackage{amsmath}
\usepackage{amssymb}
\makeatletter

\def\ps@pprintTitle{%
 \let\@oddhead\@empty
 \let\@evenhead\@empty
 \def\@oddfoot{}%
 \let\@evenfoot\@oddfoot}
 
\newcommand\footnoteref[1]{\protected@xdef\@thefnmark{\ref{#1}}\@footnotemark}
\makeatother

\usepackage[left=3.cm, right=3.0cm]{geometry} 

\usepackage{algorithm}% http://ctan.org/pkg/algorithms
\usepackage{algpseudocode}% http://ctan.org/pkg/algorithmicx

\usepackage{todonotes}

\journal{Astronomy and Computing}

\biboptions{authoryear}

\begin{document}

\begin{frontmatter}

\title{Fast Sampling from Wiener Posteriors for Image Data with Dataflow Engines}

\tnotetext[mytitlenote]{Article accepted for publication Astronomy and Computing 3 Oct 2018 }

% Group authors per affiliation:
\author{Niall Jeffrey$^{1,2}$}
\ead{niall.jeffrey.15@ucl.ac.uk}
\author{Alan F. Heavens$^{2}$, Philip D. Fortio$^{2}$}
\address{$^{1}$Department of Physics and Astronomy, University College London, Gower Place, London WC1E 6BT, UK}
\address{$^{2}$Imperial Centre for Inference and Cosmology (ICIC), Imperial College London, London SW7 2AZ, UK}

\begin{abstract}
We use Dataflow Engines (DFE) to construct an efficient Wiener filter of noisy and incomplete image data, and to quickly draw probabilistic samples of the compatible true underlying images from the Wiener posterior.  {\color{black} Dataflow computing is a powerful approach using reconfigurable hardware, which can be deeply pipelined and is intrinsically parallel. The unique Wiener-filtered image is the minimum-variance linear estimate of the true image (if the signal and noise covariances are known) and the most probable true image (if the signal and noise are Gaussian distributed). However, many images are compatible with the data with different probabilities, given by the analytic posterior probability distribution referred to as the Wiener posterior.} The DFE code {\color{black} also} draws large numbers of samples of true images from this posterior{\color{black}, which allows for further statistical analysis}. Naive computation of the Wiener-filtered image is impractical for large datasets, as it scales as $n^3$, where $n$ is the number of pixels. We use a messenger field algorithm, which is well suited to a DFE implementation, to draw samples from the Wiener posterior, that is, with the correct probability we draw samples of noiseless images that are compatible with the observed noisy image. {\color{black} The Wiener-filtered image can be obtained by a trivial modification of the algorithm.} We demonstrate a lower bound on the speed-up, from drawing $10^5$ samples of a $128^2$ image, of {\color{black} 11.3 $\pm$ 0.8} with 8 DFEs in a 1U $\rm \tt MPC$-$\rm \tt X$ box when compared with a 1U server {\color{black} presenting 32 CPU threads}. We also discuss a potential application in astronomy, to provide better dark matter maps and improved determination of the parameters of the Universe.
\end{abstract}


\end{frontmatter}

% \linenumbers

\section{Introduction}

{\color{black} Dataflow computing has recently aided the significant acceleration of many computationally-intensive and data-intensive problems. This paper discusses the use of Dataflow Engines (DFEs) for sampling realisations of noise-free images from the Wiener posterior distribution given noisy and incomplete data, with particular applicability to astronomy and cosmology.}

The Wiener filter~\citep{wiener1949extrapolation} is a useful statistical tool in many image analyses, as it is a minimum variance linear filter, and moreover the filtered data are also the \textit{maximum a posteriori} (MAP) values if the data have Gaussian signal and noise. To be more specific, if the covariance matrices of the noise and signal are known, then the Wiener filtered image has the smallest variance of any linear-filtered image. %Fast evaluation of the Wiener filter for an image and samples from the Wiener posterior probability distribution for each pixel can be extremely useful. 
Mathematically it is straightforward to write down the expression for the Wiener-filtered image, and the covariance of compatible images, but evaluation is problematic as it involves the inversion of large matrices that are in general non-diagonal. As image datasets become larger, naive Wiener methods become unfeasible (requiring approximations such as re-binning to larger pixels or assuming white noise).

By using messenger field algorithms (described in section~\ref{sec:messenger}) the Wiener image and posterior can be computed feasibly, with no need to simplify the existing algorithms.  Furthermore, the repeated operations inherent in drawing samples from the Wiener posterior lend themselves to efficient computation on DFEs, and we demonstrate that by a comparison with an implementation on multiple CPUs.

\subsection{Data model}

Although the typical applications of Wiener filters involve 2D image data, the formalism is general.  In any case, we arrange the 2D pixel data as a list, and thus describe it by a data vector $\mathbf{d}$, and the true image is similarly described by a vector $\mathbf{s}$.

Our linear data model assumes that data $\mathbf{d}$ and true signal $\mathbf{s}$ are related by

\begin{equation} \label{eq:linear}
\mathbf{d} = \mathbf{A} \mathbf{s} + \mathbf{n} \ .
\end{equation}
where $\mathbf{n}$ is random noise, and there is a known linear operator matrix $\mathbf{A}$, which in the simplest case is just the identity matrix.

\noindent The Wiener filter $\mathbf{W}$ (\citealt{wiener1949extrapolation},~\citealt{zaroubi_wiener}) is given by

\begin{equation} \label{eq:wiener_filter}
\mathbf{W} = \mathbf{S} \mathbf{A}^\dagger \big( \mathbf{A} \mathbf{S} \mathbf{A}^\dagger + \mathbf{N} \big)^{-1}  \ ,
\end{equation}

\noindent and the Wiener filtered solution, which is the minimum-variance linearly-obtained solution for the true image, is 

\begin{equation} \label{eq:wiener_solution}
\mathbf{s}_W = \mathbf{W} \mathbf{d} \ .
\end{equation}

\noindent In these equations, $\mathbf{S} = \langle \mathbf{s} \mathbf{s}^\dagger \rangle$ and $\mathbf{N} = \langle \mathbf{n} \mathbf{n}^\dagger \rangle$ are the signal and noise covariance matrices respectively, which are assumed to be known, and we have assumed that $\langle \mathbf{s} \rangle = \langle \mathbf{n} \rangle = 0$ for simplicity (this can easily be relaxed). The angle brackets indicate the expectation value, equal to the average over infinitely many realisations of the signal for ergodic fields. If, as we will assume, the pixel noise is uncorrelated, then $\mathbf{N}$ is diagonal in pixel space. In addition to pixel noise, missing data in a given pixel can be incorporated into the Wiener filter by setting the pixel noise variance to infinity.

{\color{black} As mentioned in section~\ref{section:theory}}, the Wiener filter reconstruction, $\mathbf{s}_W$, is the linear minimum variance filter for a given $\mathbf{S}$ and $\mathbf{N}$ regardless of the statistical properties of either the signal or the noise. Note that the Wiener filtered image {\color{black} variance is biased low; e.g. high intensity pixels are suppressed.} For Gaussian signal and noise, the Wiener filter additionally becomes the MAP estimate. In addition to computing the MAP estimate, for statistical purposes it is often useful to draw samples of maps, that are compatible with the data, with the appropriate probability.  These can be used for subsequent statistical analysis of the true image, such as determining the uncertainty in a given pixel. This is discussed further in section~\ref{sec:wiener_posterior}.

Calculation of the Wiener filter is challenging due to the inversion of covariance matrices, which may not be diagonal, and can become prohibitively time consuming for large images, especially when one notes that for an $N \times N$ image, the matrices are $N^2 \times N^2$ in size.

In some applications the signal is statistically {\color{black} homogeneous}, leading to a diagonal signal covariance in the Fourier/harmonic domain, which leads to a route to a solution that does not involve the inversion of large non-diagonal matrices \citep{elsner2013}.  This is not trivial, since although independent noise has a diagonal covariance matrix in pixel space, it is not diagonal in harmonic space if the dataset has varying noise variance and is thus heteroscedastic. This situation automatically arises if there are missing data, but not only in this case. Therefore, in general there is no natural basis in which both the signal and noise covariance matrices are sparse. It is possible to take advantage of the bases in which the covariance matrices are sparse by using algorithms that employ so-called ``messenger fields''~\citep{elsner2013} to convey information between harmonic and pixel space.

The messenger field class of algorithms is highly suited to a Dataflow implementation. Using reconfigurable hardware accelerators rather than CPUs helps to deal iteratively with large volumes of data. {\color{black} DFEs have recently been successfully applied to a wide range of scientific problems, including geoscience~\citep{gan_fu_mencer_luk_yang_2017}, fluid-dynamics~\citep{duben2015use}, artificial neural networks~\citep{liang2018fp}, quantum chemistry~\citep{quantum_chem}, and genomics~\citep{arram2015fpga}.}

In Section 2, we describe the Wiener filter in a Bayesian framework, and show how messenger fields are used to draw samples from the Wiener posterior probability distribution. In Section 3, we describe Dataflow computing and present our implementation of the Wiener sampler. We present the results in Section 4. In Section 5, we describe our motivation for this work as an application to upcoming large cosmology surveys.

\section{Theoretical Background} \label{section:theory}

\subsection{Wiener Posterior} \label{sec:wiener_posterior}

For the linear model of equation~\ref{eq:linear}, the Wiener filter, with $\mathbf{W}$ given by equation~\ref{eq:wiener_filter}, is a linear operator which minimises the variance

\begin{equation}
V = \langle (\mathbf{Wd} - \mathbf{s})^\dagger (\mathbf{Wd} - \mathbf{s}) \rangle \ .
\end{equation}

From a different starting point, for the Wiener posterior, we begin by assuming a Gaussian likelihood for the pixel noise\footnote{We can also argue that if only the covariance and the mean is known, the Gaussian distribution is most appropriate to assume, as it is the maximum entropy distribution.}~\citep{jasche15}:

\begin{equation}
Pr( \mathbf{d} | \mathbf{s}, \mathbf{N} ) = \frac{1}{\sqrt[]{(\mathrm{det} 2 \pi \mathbf{N})}} \mathrm{exp} \Big[ - \frac{1}{2} ( \mathbf{d} - \mathbf{A} \mathbf{s} )^\dagger \mathbf{N}^{-1} ( \mathbf{d} - \mathbf{A} \mathbf{s} ) \Big] \ .
\end{equation}

Assuming that the prior on the signal is that of a Gaussian random field,

\begin{equation}
Pr( \mathbf{s} | \mathbf{S} ) = \frac{1}{\sqrt[]{(\mathrm{det} 2 \pi \mathbf{S})}} \mathrm{exp} \Big[ - \frac{1}{2} \mathbf{s}^\dagger \mathbf{S}^{-1} \mathbf{s}  \Big] \ ,
\end{equation}

\noindent then using Bayes' theorem and the fact that $Pr(\mathbf{d} | \mathbf{S}, \mathbf{s}, \mathbf{N} ) = Pr(\mathbf{d} | \mathbf{s}, \mathbf{N} )$, the full Wiener posterior can be found:

\begin{equation} \label{eq:wienerpost}
\begin{split}
Pr( \mathbf{s} | \mathbf{S}, \mathbf{N}, \mathbf{d} ) &= \frac{  Pr( \mathbf{d} | \mathbf{s}, \mathbf{N} ) Pr( \mathbf{s} | \mathbf{S}, \mathbf{N} ) }{Pr( \mathbf{d} |  \mathbf{N} )} \\
 &= \frac{1}{\sqrt[]{(\mathrm{det} 2 \pi \mathbf{S})}} \frac{1}{\sqrt[]{(\mathrm{det} 2 \pi \mathbf{N})}}  \mathrm{exp} \Big[ - \frac{1}{2} \mathbf{s}^\dagger \mathbf{S}^{-1} \mathbf{s}  - \frac{1}{2} ( \mathbf{d} - \mathbf{A} \mathbf{s} )^\dagger \mathbf{N}^{-1} ( \mathbf{d} - \mathbf{A} \mathbf{s} ) \Big] \\
 &\propto \mathrm{exp} \Big[  - \frac{1}{2} ( \mathbf{s} - \mathbf{W} \mathbf{d} )^\dagger (  \mathbf{S}^{-1} + \mathbf{A}^\dagger \mathbf{N}^{-1} \mathbf{A} ) (\mathbf{s} - \mathbf{W} \mathbf{d}) \Big] \ .
\end{split}
\end{equation}

\noindent Here we see that the \textit{maximum a posteriori} (MAP) solution is indeed that of the Wiener reconstruction, $\mathbf{s}=\mathbf{W} \mathbf{d}$.

If we can handle the large matrices, realisations of the true underlying signal image $\mathbf{s}$ can be drawn from the posterior distribution $Pr( \mathbf{s} | \mathbf{S}, \mathbf{d} )$. The expected mean of these samples is the Wiener-filtered image. Drawing samples from the Wiener posterior clearly also suffers from the need to invert large matrices with no natural sparse basis.

Progress can be made for signal images with statistical properties that are independent of pixel position $\mathbf{x}$ {\color{black}(i.e. statistically homogeneous signals)}, for in this case, the Fourier transform of the image $\mathbf{s}_\mathbf{x}$,
\begin{equation}
\tilde{{s}}_\mathbf{k} = \sum_\mathbf{x} s_\mathbf{x} e^{- i \mathbf{k} \cdot \mathbf{x}} \ 
\label{FT}
\end{equation}
has a diagonal covariance matrix,

\begin{equation} \label{eq:power_spectrum}
\langle\tilde{{s}}_\mathbf{k}\tilde{{s}}^*_\mathbf{k'}\rangle = P(k) \delta_{\mathbf{k}\mathbf{k'}} 
\end{equation}

\noindent and $\delta_{\mathbf{k}\mathbf{k'}}$ is a Kronecker delta for the discrete 2D wavenumbers $\mathbf{k}$ and $\mathbf{k'}$, and $P(k)$ is the power spectrum, which depends only on the magnitude $k\equiv |\mathbf{k}|$.  The covariance matrix $\mathbf{S}$ for the signal is diagonal, with entries given by the appropriate $P(k)$.

\subsection{Messenger Fields} \label{sec:messenger}

The messenger field approach splits the problem into two, performing some operations in harmonic space and some in pixel space, transferring the information using an extra field, $\mathbf{t}$, called the messenger field, whose covariance matrix is diagonal in both spaces. The method takes advantage of the diagonal signal covariance matrix in harmonic space and the diagonal noise covariance matrix in pixel space, such that no matrices need to be inverted in a basis in which they are dense.

This field is defined to have zero mean and a covariance matrix proportional to the identity matrix, $\langle \mathbf{t} \mathbf{t}^\dagger \rangle \propto \mathbf{I}$, which will always be diagonal in both harmonic and pixel spaces. The Markov Chain Monte Carlo (MCMC) algorithm used in~\cite{jasche15} is a method which uses the messenger field to draw samples from the Wiener posterior, without inversion of non-diagonal covariance matrices, requiring instead repeated Fourier transforms and inverse Fourier transforms. The algorithm is presented in Algorithm~\ref{alg:messenger}. In the limit of large numbers of iterations, this unconditionally converges to drawing samples from the desired distribution. 

A sufficient number of samples from the Wiener posterior probability distribution can characterize the statistical properties of the underlying signal given some data. 

\begin{algorithm}
\color{black}
\caption{Messenger Field Wiener Sampler: an iterative method to draw sample signal images from a Wiener posterior distribution using messenger fields~\citep{jasche15} \label{alg:messenger}}
\begin{algorithmic}[1]
\Procedure{Sampler}{}
\State \hspace{0.0cm} \rm{for} $t_i$ in $\mathbf{t}$: \label{alg:start}
\State \hspace{0.5cm} $t_i = \mu_i^t + \sqrt[]{(\sigma_i^t)^2} \, G(0,1)$
\State \hspace{0.0cm} $\hat{\mathbf{t}} = \mathcal{F}_{2D} (\mathbf{t})$ 
\State \hspace{0.0cm} for $\hat{s}_k$ in $\hat{\mathbf{s}}$:
\State \hspace{0.5cm} $\hat{s}_k = \mu_k^s + \sqrt[]{(\sigma_k^{\hat{s}})^2} \, G(0,1)$
\State \hspace{0.cm} $\mathbf{s} = \mathcal{F}^{-1}_{2D} (\hat{\mathbf{s}})$
\State \hspace{0.cm} Return $\mathbf{s}$
\State \hspace{0.cm} GOTO line~\ref{alg:start}
\EndProcedure
\end{algorithmic}
Definitions: \\ \vspace{-0.8cm}
\begin{itemize}
\item $\mu_i^t = \frac{T_i}{T_i A_i^2+ \bar{N}_i} A_i d_i + \frac{\bar{N}_i}{T_i A_i^2+ \bar{N}_i} s_i$ if $A_i^2 > 0$ \\  \vspace{-0.8cm}
\item $(\sigma_i^t)^2 = \frac{T_i \bar{N}_i}{T_i A_i^2+ \bar{N}_i} $  if $A_i^2 > 0$ \\ \vspace{-0.8cm}
\item $\mu_k^{\hat{s}} = \frac{\hat{S}_k}{\hat{S}_k + \hat{T}_k} \hat{t}_k$ \\ \vspace{-0.8cm}
\item $(\sigma_k^{\hat{t}})^2 = \frac{\hat{S}_k \hat{T}_k}{\hat{S}_k + \hat{T}_k}$ \\  \vspace{-0.8cm}
\item $\mathbf{T} = \mathrm{min} \big( (\mathbf{A}^{-1})^\dagger  \mathbf{N}  (\mathbf{A}^{-1})  \big) \mathbf{I} $  \\ \vspace{-0.8cm}
\item $\bar{\mathbf{N}} = \mathbf{N} - \mathbf{A}^\dagger \mathbf{T} \mathbf{A} $ \\ \vspace{-0.8cm}
\item $G(0,1)$ is a zero-mean Gaussian random variate with unit variance. \\ \vspace{-0.8cm}
\item $\mathcal{F}_{2D}$ is the 2D Fourier transform and $\mathcal{F}^{-1}_{2D}$ its inverse.
\end{itemize}
\end{algorithm}

By replacing the random variates in Algorithm~\ref{alg:messenger} with zero ($G(0,1) \rightarrow 0$), the iteration outputs converge to the (unsampled) Wiener filter reconstruction (equation~\ref{eq:wiener_filter}). This was first described by~\cite{elsner2013} in the first use of messenger fields. With this small change the code provides Wiener-filtered images, rather than samples from the Wiener posterior. 

For each calculation of a Wiener filter or sample drawn from the Wiener posterior, $\mathcal{O}(n^3)$ operations are required for a length $n$ data vector using a naive approach. Using the messenger field algorithm, this reduces to $\mathcal{O}(n \log n)$ for covariance matrices that are diagonal in their respective domains. The naive approach is bottlenecked by the matrix inversion and the messenger field approach is bottlenecked by the harmonic/Fourier transform.

\section{DFE Implementation}

\subsection{DFE System}

The standard computing paradigm in the present day still follows the outline of the von Neumann model, often called Control Flow. In a standard setup a Central Processing Unit (CPU) carries out computational operations with data and instructions provided by memory, usually Random Access Memory (RAM). Data and instructions are iteratively passed between memory and CPU. 

Dataflow Engines (DFEs) use reconfigurable hardware rather than CPUs to represent a static description of an algorithm with deep hardware pipelines consisting of a series of standard arithmetic and logic operations. DFEs do not need to continually get new instructions from the memory~\citep{pell2013maximum}. They are therefore intrinsically parallel. The Wiener sampling problem described above has a high volume of data with highly deterministic computation (few ``if'' statements), so is well suited to DFEs. 

Unlike standard CPU-based High Performance Computing (HPC) platforms, DFEs can be reconfigured on occasion to the need of a given algorithm or dataset. For the cost of an initial build time ($\mathcal{O}(\mathrm{hours})$), the speed and efficiency at runtime is improved. These systems allow greater flexibility with memory, data type, and clock frequency.

{\color{black}For example, higher clock frequencies can lead to shorter run time of the compute kernels instantiated on the DFE. This can yield faster execution if the algorithm is compute bound. However, for higher clock frequencies it becomes more difficult to build the reconfiguration bitstream, so the clock frequency can be chosen optimally for a given algorithm.}

The CPU code for managing a DFE can be written in $\rm \tt C$ or $\rm \tt C$++ and runs on a host (a traditional control flow machine). {\color{black}For the DFE, the software is written in Java-like code, which is compiled into the reconfiguration file for the hardware chip. This turns the DFE into a problem-specific hardware accelerator.} 

Once reconfigured, the DFE accepts data streaming and compute action calls launched by the host CPU code. {\color{black}  A single DFE is a PCIe card that can either be available locally on a CPU server or be mounted in a Maxeler $\tt MPC$-$\rm \tt X$: a CPU-free 1U server appliance hosting up to 8 DFEs, which is connected to host CPU servers by an infiniband network. Each DFE carries a chip with large amount of reconfigurable logic and on-chip resources (e.g. a Field-Programmable Gate Array, FPGA) with up to 96\,GB of on-board DRAM storage. The MAX4 generation cards available to the authors are MAX4 Maia DFEs with Altera Stratix V FPGA and 48\,GB of DRAM.} Integration of CPU and DFE codes is done by the dedicated compiler as described in~\cite{krt2015}.

Dataflow Engines allow user-friendly control over the features of the underlying hardware, so the hardware description can be optimally designed and built for the algorithm at hand. This can lead to large speed-ups at runtime compared to the same algorithm's implementation on a comparative CPU platform. 

Time to complete a task is also only one metric of performance among other metrics. Lower clock frequencies mean that DFEs use less power than conventional CPU machines~\citep{gan_fu_mencer_luk_yang_2017}. {\color{black} Usually FPGAs use an order of magnitude less power than CPUs~\citep{liang2018fp}}. In many applications, it is therefore more cost efficient to use DFEs as it allows more science per Watt.

{\color{black} Another commonly used and increasingly popular alternative to CPU hardware are graphics processing units (GPU), which gain acceleration for vectorized problems using ``single instruction, multiple data'' (SIMD) architectures and high-clock frequency~\citep{liang2018fp}. However, they are disadvantaged by their high energy cost. CPUs are more efficient than GPUs, and, as discussed, FPGAs are in turn more efficient than CPUs. GPUs additionally do not benefit from the flexibility that allows reconfigurable DFEs to tailor to a specific algorithm. Their hardware cannot be optimally designed for a given problem.}

\subsection{Implementation} \label{sec:implementation}

\begin{figure}[H]
\centering
\includegraphics[width=0.65\textwidth]{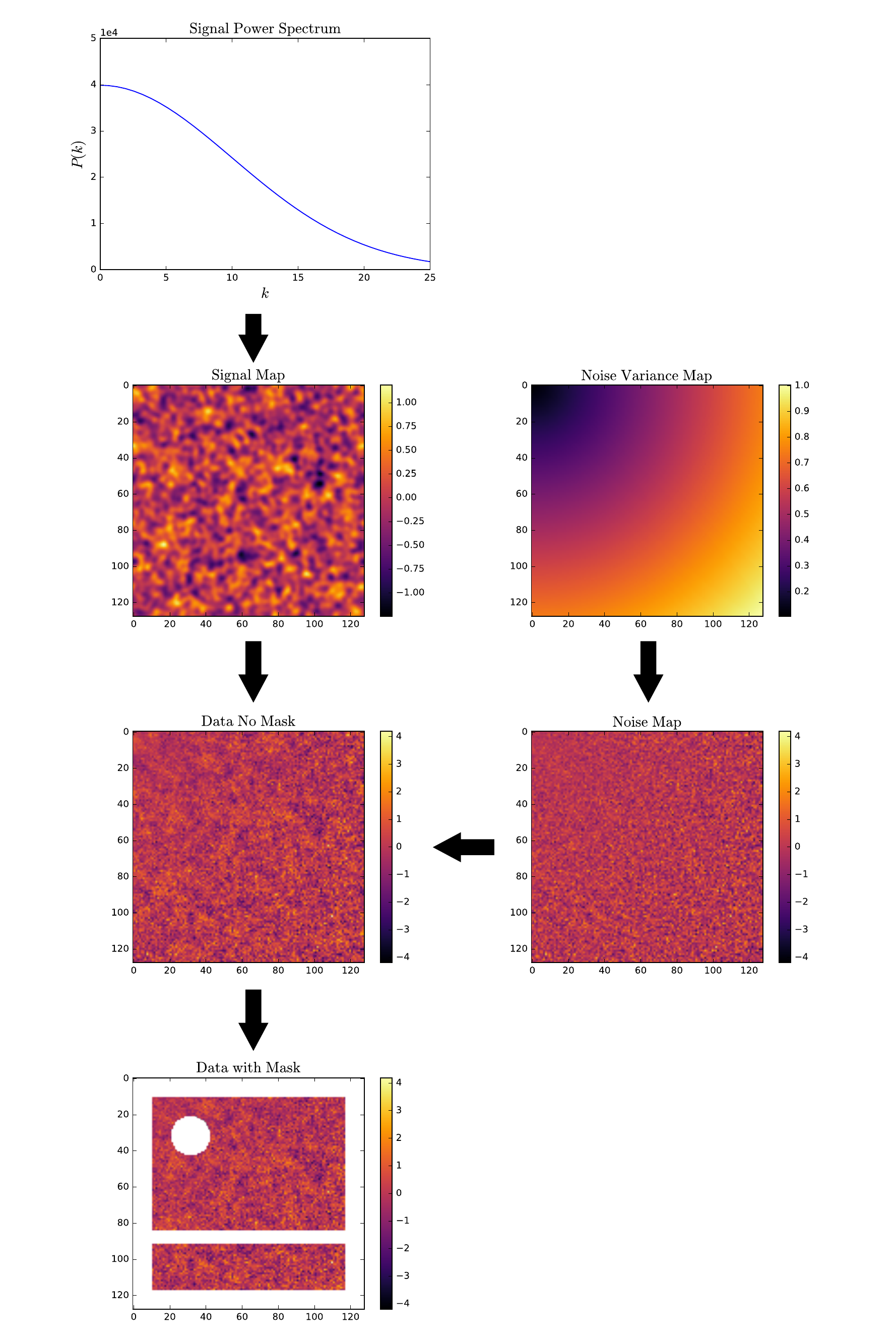}
\caption{\label{fig:data_generation} This figure shows our data model, and gives an example realisation of a simulated dataset. We begin with a \textbf{\textit{Signal Power Spectrum}}, $P(k)$, from which we generate a real, Gaussian random field as a \textbf{\textit{Signal Map}}. We then take a \textbf{\textit{Noise Variance Map}}, whose values vary across the data, from which we generate a \textbf{\textit{Noise Map}} of Gaussian, independent pixel noise. The noise is added to the signal to generate the \textbf{\textit{Data No Mask}}. We mask pixels representing missing data in \textbf{\textit{Data with Mask}}.}
\end{figure}

We show the steps taken to generate a typical simulated dataset with the desired properties in figure~\ref{fig:data_generation}.  To simulate underlying signals $\mathbf{s}$, we generate realisations of square, two-dimensional images, which are in this case real, zero-mean, Gaussian random fields with known power spectra. The real and imaginary parts of $s_\mathbf{k}$ are each drawn randomly from Gaussian distributions with variance $P(k)/2$, and reality of the signal is enforced by $s^*_\mathbf{k} = s_\mathbf{-k}$. 
We simulate square signal maps with $128^2$ pixels.

The image, and therefore the vectors $\mathbf{k}$ and $\mathbf{x}$, are two dimensional, so the transforms employs a 2D DFT. In practice the fast Fourier transform (FFT) algorithm~\citep{cooley1965algorithm} is used to evaluate the coefficients.

The datasets are generated according to the linear model of equation~\ref{eq:linear}. For simplicity, we do not apply the linear operator (setting $\mathbf{A} = \mathbf{I}$), though this could be included for a given application. The first panel of figure~\ref{fig:data_generation} shows an initial power spectrum, $P(k)$, from which we generate our real, Gaussian field as the signal map. 

The noise is independent between pixels and is drawn from a Gaussian distribution where the noise variance varies across the data. We assume that the noise variance is known. We mask some of the pixels to represent missing data. The Wiener filter and the Wiener posterior treat the missing data as a special case of infinite noise. Infinite noise variance, in the region of the missing data, is set to be $10^8$, as an effective infinity. 

\begin{figure}[H]
\centering
\includegraphics[width=0.6\textwidth]{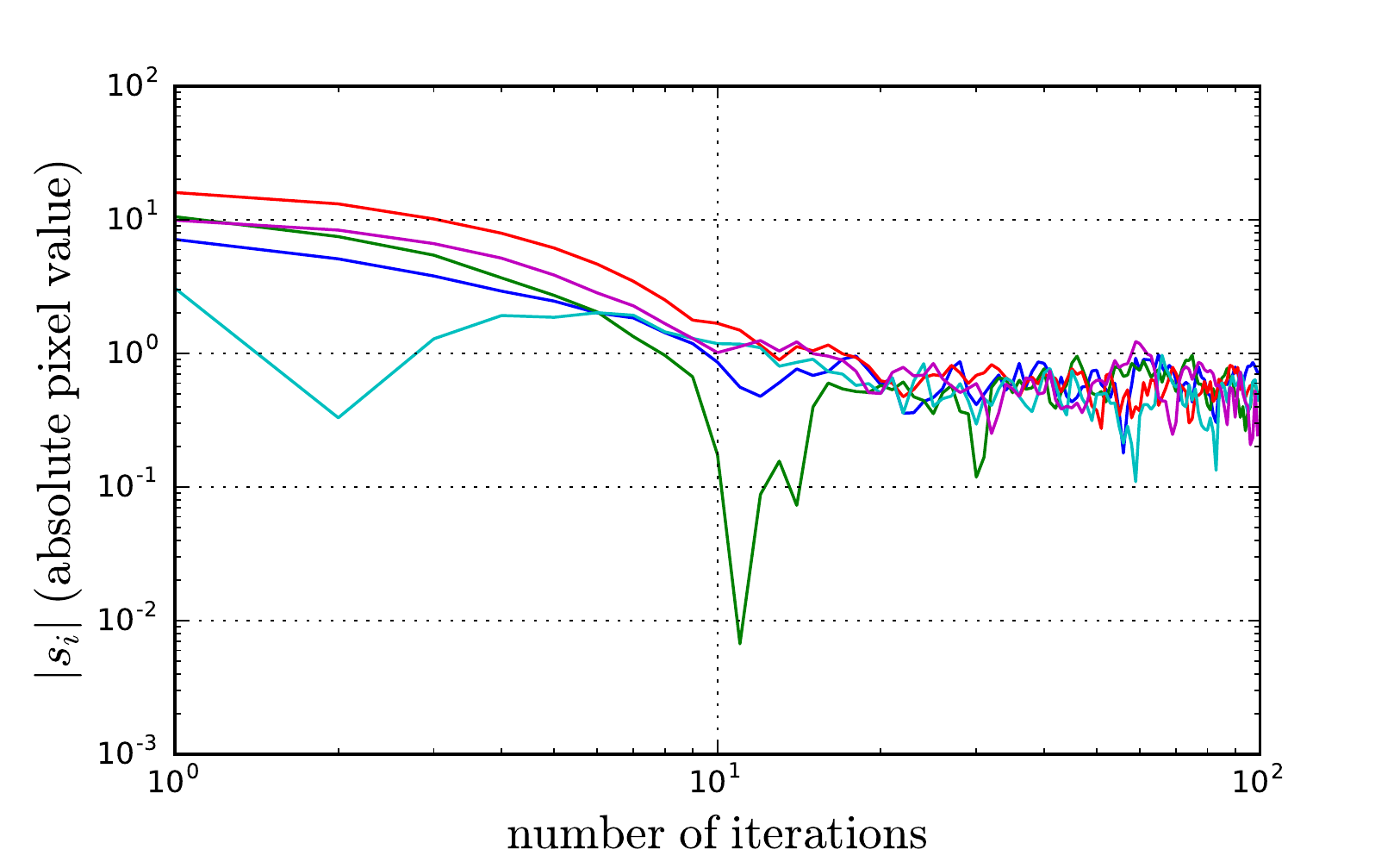}
\caption{\label{fig:convergence} The absolute value of the same pixel at each iteration in 5 independent MCMC chains of the messenger field algorithm. The initial value of each pixel is different, to show convergence after a sufficient number of iterations.}
\end{figure}

On both CPU and DFE, we implement the messenger field algorithm (Algorithm~\ref{alg:messenger}) to draw samples of signal from the Wiener posterior (equation~\ref{eq:wienerpost}), using 5 different datasets at each iteration. This reflects~\cite{alsing17}, where multiple chains were run in parallel to test convergence. In figure~\ref{fig:convergence}, the value of the same pixel in 5 independent chains with different initial values can be shown to converge after a sufficient number of iterations. The period during which the chains have not converged is known as burn-in, and using these samples reduces the influence of the initial starting point. Subsequent points are not converged immediately, therefore it is essential to have multiple chains, to check convergence and improve statistics.

On DFEs it is possible to instantiate fixed point and IEEE-like floating point arithmetic units of an arbitrarily chosen bitwidth, with greater flexibility beyond the standard options of single- or double-precision floating-point. Reducing the bitwidth of the number representation results in less reconfigurable logic spent on single arithmetic operations. This allows the user to instantiate more arithmetic units to fit the budget of reconfigurable space available on the chip, which may be used to implement more complex logic, or to replicate the computational pipeline; the latter reduces time to solution due to increased parallelism, but at the cost of reduced precision. In the implementation presented in this paper, we use single-precision floating point format on both the DFE and on the CPU, to compare more easily the results.

The CPU code, written in $\rm \tt C$, uses a Box-Muller transformation to generate pairs of normally-distributed random variates for use in the algorithm. This custom-written implementation was shown to be consistently faster than the $\rm \tt std\ C{++}$ Gaussian random number generator in unit tests. Our implementation is slightly faster as we only ever generate one pair of zero mean and unit variance Gaussian random numbers at each iteration. The DFE uses the Gaussian random number generator from a dedicated dataflow library~\footnote{\label{note1}\color{black} MaxPower (\url{maxeler.com/mymaxeler} requires Maxeler account)}. This small difference changes the overall time measurement little, as the fraction of time spent generating random numbers is small in this algorithm. 

The 2D FFT from the $\rm \tt FFTW3$ package~\citep{FFTW05} was used for the CPU code, {\color{black}optimised with Advanced Vector Extensions (AVX2) available on the CPU hardware (see section~\ref{sec:intel})}. A dedicated dataflow FFT library was used for the DFE\footnoteref{note1}.

\section{Results}

\subsection{Wiener posterior properties}

\begin{figure}[H]
\centering
\hspace*{-0.25in}
\includegraphics[width=1.0\textwidth]{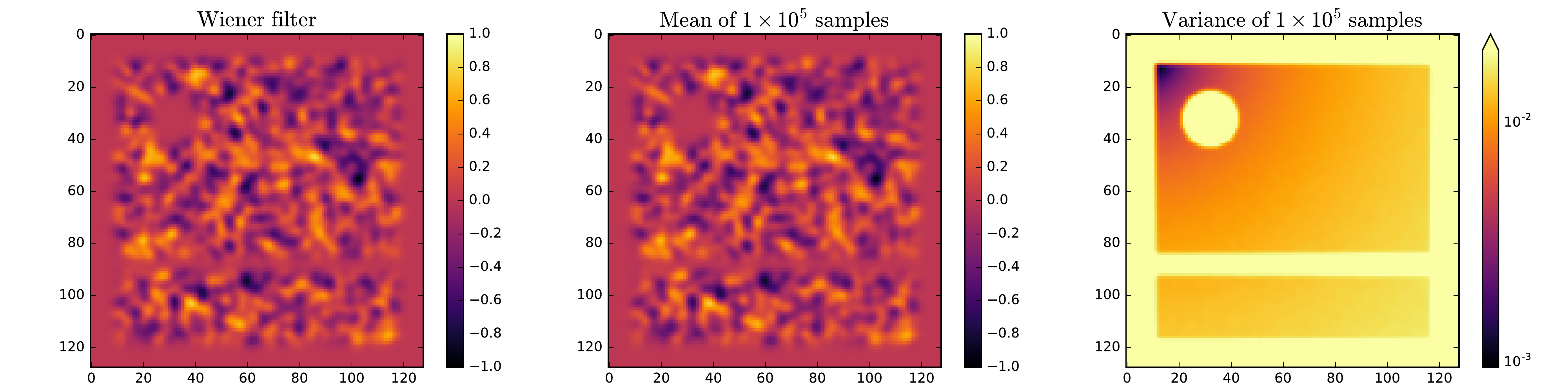}
\caption{\label{fig:mean_and_variance} The input data, signal power spectrum and noise variance are shown in figure~\ref{fig:data_generation}. \textit{Left panel}: the left panel is the Wiener filter solution, where the random variates are not included in Algorithm~\ref{alg:messenger}. \textit{Centre panel}: Mean of $10^5$ samples from the Wiener posterior distribution evaluated using Algorithm~\ref{alg:messenger}. \textit{Right panel}: Variance of the same set of samples.}
\end{figure}

As described in section~\ref{sec:messenger}, samples from the Wiener posterior sampling algorithm without the random variates converge to the Wiener filter solution (equation~\ref{eq:wiener_filter}). In the left panel of figure~\ref{fig:mean_and_variance} the Wiener filter reconstruction from this method is shown for the data generated in figure~\ref{fig:data_generation}. By doing this we tested that the CPU and DFE outputs are identical up to computational precision. 

A~second test also provided the DFE with a vector of random Gaussian variates, shared with the CPU, where the output samples from the Wiener posterior were shown to be the same within computational precision.

In the centre panel of figure~\ref{fig:mean_and_variance}, the mean of the $10^5$ samples from the Wiener posterior can be seen. By comparing to the Wiener filtered image in the left panel, one can see that the Wiener filtered solution is indistinguishable from the mean of the samples from the Wiener posterior, as expected. Due to sample variance, the mean of samples from Wiener posterior is not exactly equal to the Wiener filter, though for an infinitely large number of samples it would be. 

In figure~\ref{fig:mean_and_variance}, the variance of the same $10^5$ samples can be seen in the right panel. The variance in the region of missing data is high, as expected, but constrained by the signal covariance. The structure of the variance of the samples matches the structure of the noise variance map (see figure~\ref{fig:data_generation}) as expected.

By drawing sufficient samples from the full posterior probability distribution, the code can characterise it very well, not just providing its mean and covariance. 
%An example use of the samples within a hierarchical Bayesian model for problems in cosmology are described in section~\ref{sec:cosmology}.

\subsection{CPU vs. DFE, Speed} \label{sec:intel}

We compare the speed of the CPU+DFE implementation of the Wiener sampler (Algorithm~\ref{alg:messenger}) to the pure CPU implementation. Both were run on an{\color{black} $\ \rm \tt Intel(R)\ Xeon(R)\ E5-2650\ v2\ @\ 2.60GHz$ server (2 sockets, 8 dual-thread cores per socket) presenting 32 CPU threads, which is} connected to a $\rm \tt MPC$-$\rm \tt X$ node at the STFC Hartree Centre. A single $\rm \tt MPC$-$\rm \tt X$ box contains $8\ \rm \tt MAX4\ (Maia)$ DFEs. The clock frequency for the DFE implementation was chosen to be $200\ \rm MHz$.

\begin{figure}[H]
\centering
\includegraphics[width=0.8\textwidth]{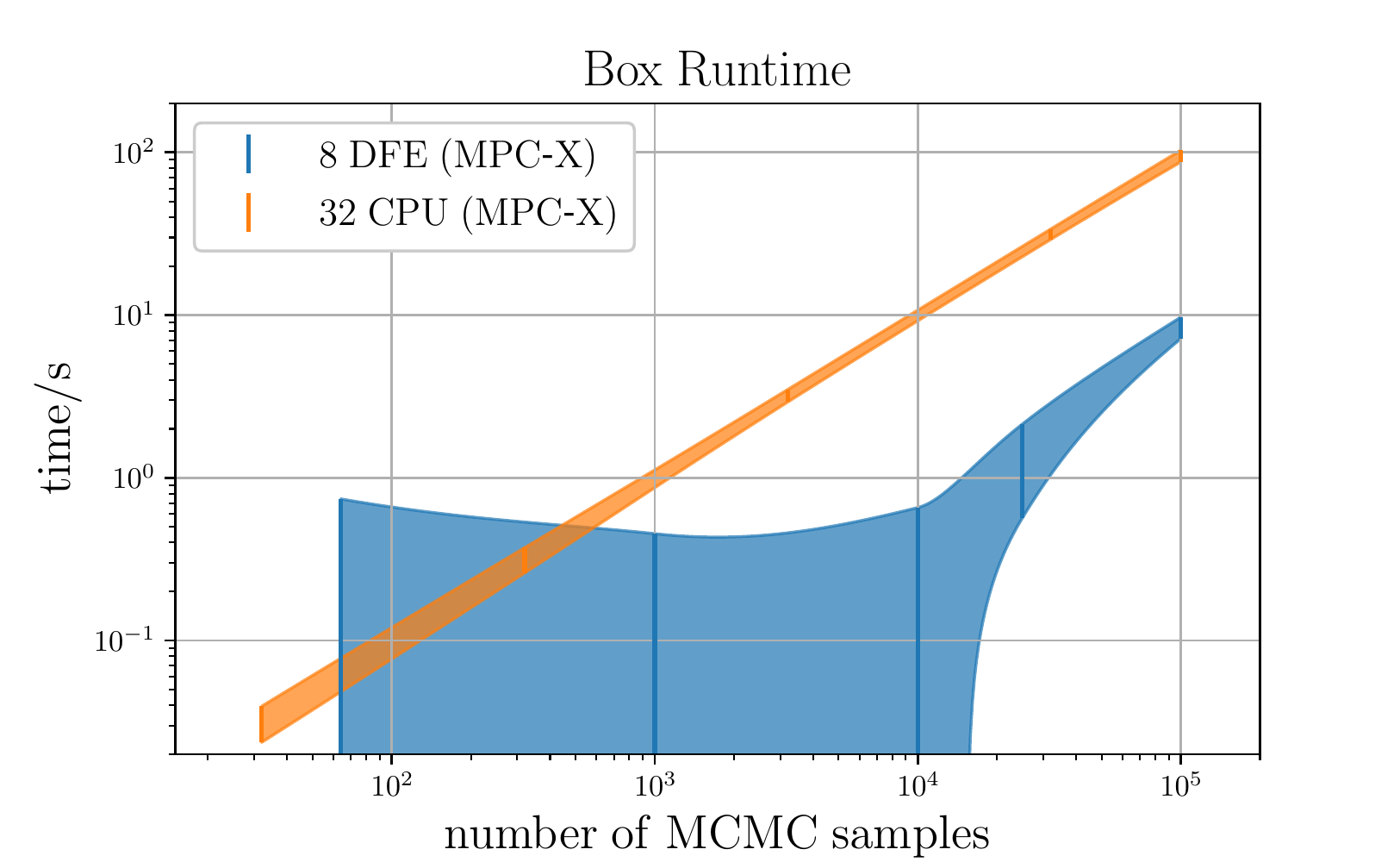}
\caption{\label{fig:times} Time taken to run the sampling algorithm for a given number of iterations, where each iteration returns a sample from each of the 5 chains. {\color{black}Each data point is the mean of 10 runs (with the DFE data overhead removed), and $1 \sigma$ error bars with~\citeauthor{akima1970new}~(\citeyear{akima1970new}) interpolation for the error envelope. 32 CPU threads vs 8 DFEs, run in parallel on $\rm \tt MPC$-$\rm \tt X$. }}
\end{figure} 

The sampler was timed for increasing number of iterations on both the CPU and the DFE, up to $10^5$ iterations. Samples of 5 images are returned at each iteration. The time was measured from the CPU from the start to the end of the algorithm's execution. At each number of iterations, the algorithm was repeated 10 times and the measured times averaged.

Figure~\ref{fig:times} shows the time to perform the algorithm for a 1U $\rm \tt MPC$-$\rm \tt X$ with 8 DFEs against a 1U server {\color{black} presenting 32 CPU threads}. The time as a function of number of iterations is linear for both CPU and DFE. {\color{black}The DFE has an initial overhead (with an average of $4.0$ seconds) as the data is loaded onto the hardware, which is removed from the DFE time. Errors are obtained from 10 runs of the code. For the low run times, the DFE times have larger error-bars than the CPU, due to larger variance in the DFE data loading time; the relative effect of this decreases with longer running times.}

In order to parallelise the problem, we run independent MCMC chains. We measure the time to generate a given number of samples by running the MCMC on 32 memory independent CPU threads for the CPU-only code. For the CPU+DFE code, we measure the time to generate a given number of samples on 8 DFEs by splitting the work across 8 CPU threads.

In this work we have one CPU thread orchestrating one DFE. In future work we need to support an $N$:$1$ ratio of $N$ CPU threads served by a single DFE. This will help to utilise all the CPU computational capacity as well as all DFEs. Therefore, the speed-up of the CPU+DFE implementation in this work is a lower bound --- here there is scope for considerable further acceleration.

From each independent MCMC chain some number of initial samples are unusable due to burn-in and are discarded. As each MCMC chain (run in parallel CPU threads) must discard the same number of initial samples, running 32 chains gives 32 times more unusable samples than  a single chains with the same number of iterations. The 32 parallel CPUs will therefore have to discard four times more samples than 8 parallel DFE-accelerated CPU threads due to burn-in. This is also a reason why the time measurement from this $\rm \tt MPC$-$\rm \tt X$ box parallel test should be interpreted as a lower bound on the potential speed up from DFEs. 

We measure the lower bound on the parallel DFE speed-up to be {\color{black}11.3 $\pm$ 0.8}, where we have again used 10 time measurements to estimate the error.

\section{Potential Applications to Cosmology} \label{sec:cosmology}

In this section we discuss some potential use cases in cosmology, although the algorithm and implementation are general and could be used in a number of contexts.

\subsection{Power Spectrum Inference}

A common problem is to extract information from the power spectrum, $P(k)$, of an underlying field, $\mathbf{s}$, as defined in equation~\ref{eq:power_spectrum}, and an extension of the DFE code can allow this.  For a zero-mean Gaussian random field, the power spectrum contains all the statistical information that defines the field. The specific aim for power spectrum inference is to calculate the posterior probability distribution of the power spectrum given a set of data. 

The standard model of cosmology predicts that the density field of the early universe will be a Gaussian random field, which persists for large cosmological scales in the late universe. Estimating the power spectrum is therefore a standard tool in many cosmological analyses with different data sets, including the early universe through Cosmic Microwave Background (CMB) radiation data~\citep{ade2016planck}. A posterior probability distribution of the power spectrum of the density field in turn leads to posterior probability distributions for the cosmological parameters. These usually include, but are not limited to: the matter density $\Omega_m$, the Dark Energy density $\Omega_\Lambda$, the Dark Energy equation of state parameter $w$, and the Hubble parameter $H_0$.

The Bayesian hierarchical inference models described by \cite{jasche15}, \cite{alsing16}, and \cite{alsing17} to infer the posterior distributions of either the power spectrum or cosmological parameters in addition to samples from the field; see Fig.\ref{fig:bayesian_network}. With Gibbs sampling, samples of both the power spectrum and the image are drawn, keeping the other temporarily fixed. For a given power spectrum, large numbers of samples from the posterior probability distribution of the underlying signal $\mathbf{s}$ can be drawn efficiently using Dataflow Engines, leading to better constraints on the power spectrum and hence on cosmological parameters.

\subsection{Cosmological Mass Mapping}

Due to the local curvature of spacetime by the matter, images of distant galaxies are deformed by the inhomogeneous matter distribution along the line of sight. This is called gravitational lensing. Any matter can contribute to the lensing effect, making it a direct probe of non-visible dark matter~\citep{KS93}. Reconstructing this density field facilitates the study of the dark matter physics, its relationship with visible matter, and can provide novel approaches to extract additional cosmological information.

The model describing this process when in the linear regime, known as weak gravitational lensing, is fully described by our linear model in equation~\ref{eq:linear}. The data $\mathbf{d}$ are images where the pixel values are the mean of galaxy shapes within that pixel. The signal $\mathbf{s}$ is a weighted, projected density field\footnote{The weighted, projected density field in mass mapping is called convergence and is denoted by $\kappa$.} in the foreground of the observed galaxies. The pixel noise, due to the intrinsic random galaxy shapes, is approximately Gaussian. The density field in the late universe on large cosmological scales is also approximately Gaussian. 

From data with these properties, the large-scale density field from weak lensing shape measurements can be principally recovered with a Wiener filter. In~\cite{des_wiener_mass_map} the messenger field Wiener filter algorithm is applied to Dark Energy Survey gravitational lensing data to generate a mass map image of the underlying density field. The Wiener filter method has also long been an established tool for reconstructing the underlying density field using only galaxy positions~\citep{lahav_wiener}, rather than using lensing data. Obtaining a large number of samples of the Wiener distribution, as is described in this work, then gives a posterior probability distribution of the density field in each pixel.

\subsection{Future Data Requirements for Cosmology}

\begin{figure}
\centering
\includegraphics[width=1.0\textwidth]{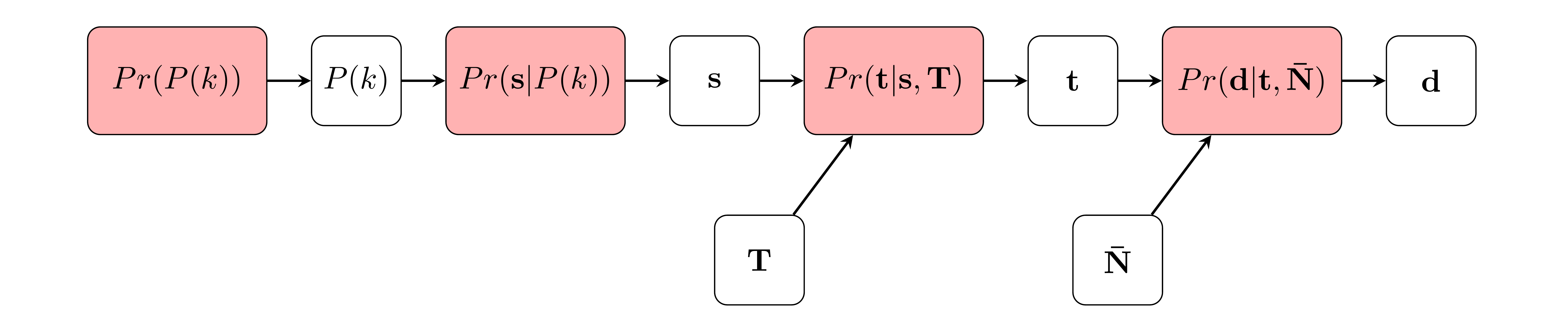}
\caption{\label{fig:bayesian_network} The Bayesian hierarchical forward model as described by~\cite{alsing17} for signal image, $\mathbf{s}$, and power spectrum, $P(k)$, inference using the messenger field, $\mathbf{t}$. The work described in this paper uses Dataflow Engines can focus on the nodes of this network that do not include the power spectrum: the power spectrum is assumed and kept constant, and samples of the signal image are drawn.}
\end{figure}

With current cosmic shear data\footnote{Cosmic shear is the spin-2 complex field manifested as the coherent distortion of galaxy shapes due to gravitational lensing. It is a function of a linear projection into 2D of the 3D density field.},~\cite{alsing17} were able to use CPUs to generate samples from the posterior probability distributions of the underlying cosmic shear signal images and the power spectrum, using the Bayesian hierarchical model shown in Fig.~\ref{fig:bayesian_network}. $10$ chains were run in parallel to a length of $10^5$ samples.

Current and future cosmic shear surveys DES~\citep{desy1}, LSST~\citep{lsst2012}, and Euclid~\citep{euclid2016}) expect orders of magnitude of increase in data volume. The European Space Agency project Euclid expects to observe over $10^9$ galaxies usable for cosmic shear, compared to $\sim 3 \times 10^6$ with the CFHTLenS data used by~\cite{alsing17}. This leap in data size requires novel computational approaches to previously tractable problems. Here, Dataflow Engines can provide a solution.  

\section{Discussion}

We have demonstrated a speed-up of at least {\color{black}11.3 $\pm$ 0.8} for generating $10^5$ samples of the Wiener posterior of possible images compatible with an observed noisy image of $128^2$ pixels, using 8 DFEs in a 1U $\rm \tt MPC$-$\rm \tt X$ box and comparing with a 1U server {\color{black}presenting} 32 CPU threads.

Future extensions could be to include the full Bayesian hierarchical model shown in figure~\ref{fig:bayesian_network}, to further exploit the increased speed afforded to us by the Dataflow approach. This would lead to better constraints on the inferred cosmological parameters through samples of the power spectrum.

For data requirements of future cosmological surveys it would be useful to Wiener filter and draw samples of the Wiener posterior from data which have more than of $128^2$ pixels per image. The image size in this work is constrained by the size of an FFT problem that fits within the fast FMEM on-chip memory ($\sim$6MB). We expect that future versions of the dataflow FFT library will provide the option to use off-chip memory (48GB) as an FFT buffer. We could then expect to be able to Fourier transform images of size $2^{15} \times 2^{15}$. This would increase the scientific applicability of a single DFE dramatically.

Implementing large scale Bayesian methods for cosmological parameter estimation on Dataflow Engines is a promising solution to the problem of increasingly large datasets from future surveys. This implementation of a Wiener sampler has broad application for inference or de-noising from any images or dataset with similar properties to those described here. 

\vspace{2cm} \noindent The authors plan to make the code public on \url{appgallery.maxeler.com} soon.

\section*{Acknowledgements}
We thank Pavel Burovskiy, Vitali Averbukh, Georgi Gaydadjiev and Edward Edmondson for useful discussions and comments. NJ acknowledges support from the UK Science and Technology Research Council (STFC) Grant No. ST/M001334/1. We acknowledge use of Hartree Centre resources in this work. The STFC Hartree Centre is a research collaboratory in association with IBM providing High Performance Computing platforms funded by the UK's investment in e-Infrastructure. The Centre aims to develop and demonstrate next generation software, optimised to take advantage of the move towards exa-scale computing.

\section*{References}

\bibliography{mybibfile}

\end{document}